\documentclass[preprint,authoryear,5p]{elsarticle}

\usepackage{graphicx}

\usepackage{amssymb}

\usepackage{hyperref}

\usepackage[final]{changes}
\usepackage{tabularx}

\journal{Advances in Space Research}

\begin{document}

\begin{frontmatter}



\title{First results from the LUCID-Timepix spacecraft payload onboard the TechDemoSat-1 satellite in Low Earth Orbit}


\author[iris,kent]{Will Furnell\corref{cor}}
\cortext[cor]{Corresponding author}
\ead{mail@willfurnell.com}


\author[iris,slbs]{Abhishek Shenoy}
\author[iris,slbs]{Elliot Fox}
\author[iris,oxford]{Peter Hatfield}
\address[iris]{The Institute for Research in Schools, Wellcome Wolfson Building, 165 Queen's Gate, London, SW7 5HD, UK}
\address[kent]{School of Computing, University of Kent, Canterbury, Kent, CT2 7NF, UK}
\address[slbs]{The Langton Star Centre, Langton Lane, Nackington Road, Canterbury, Kent, CT4 7AS, UK}
\address[oxford]{Clarendon Laboratory, University of Oxford, Parks Road, Oxford OX1 3PU, UK}

\begin{abstract}

The Langton Ultimate Cosmic ray Intensity Detector (LUCID) is a payload onboard the satellite TechDemoSat-1, used to study the radiation environment in Low Earth Orbit ($\sim$635km). LUCID operated from 2014 to 2017, collecting over 2.1 million frames of radiation data from its five Timepix detectors on board. LUCID is one of the first uses of the Timepix detector technology in open space, with the data providing useful insight into the performance of this technology in new environments. It provides high-sensitivity imaging measurements of the mixed radiation field, with a wide dynamic range in terms of spectral response, particle type and direction. The data has been analysed using computing resources provided by GridPP, with a new machine learning algorithm that uses the Tensorflow framework. This algorithm provides a new approach to processing Medipix data, using a training set of human labelled tracks, providing greater particle classification accuracy than other algorithms. For managing the LUCID data, we have developed an online platform called Timepix Analysis Platform at School (TAPAS). This provides a swift and simple way for users to analyse data that they collect using Timepix detectors from both LUCID and other experiments. We also present some possible future uses of the LUCID data and Medipix detectors in space.

\end{abstract} 


\end{frontmatter}

\parindent=0.5 cm


\section{Introduction}

Cosmic radiation consists of high energy particles produced by a variety of extra-terrestrial sources. In general, cosmic radiation falls into three categories depending on their source; Galactic Cosmic Rays (GCRs) which originate outside the solar system, Solar Energetic Particles (SEPs) which come from the sun, and charged particles that are trapped by the Earth's magnetic field.\\

When the particles are detected directly they are known as primary particles; particles produced by an interaction between a primary particle and some obstructing medium (e.g. hitting the Earth's atmosphere) are known as secondary particles. Primary cosmic rays are made up of a variety of particles (protons, electrons, gammas, light nuclei); an even wider range of particles are typically produced in secondary particle showers, including neutrons, Minimum Ionising Particles (MIPS), which are usually muons, and pions. Primary cosmic rays span a vast range of energies ($\sim10^4-10^{20}\mathrm{eV}$ - in contrast the maximum energy reached by the Large Hadron Collider is $\sim10^{13}\mathrm{eV}$, CMS Collaboration 2017) from a large range of sources, from solar (typically $E<10^9\mathrm{eV}$, including Solar Energetic Particles) to within the Milky Way (typically, $10^9\mathrm{eV}<E<10^{15}\mathrm{eV}$ Galactic Cosmic Rays), to extragalactic (typically $10^{15}\mathrm{eV}<E$), with the source of the highest energy particles ($10^{19}\mathrm{eV}<E$) still heavily debated (see \citealp{Greisen1966EndSpectrum}, \citealp{PierreAugerCollaboration2017ObservationEV.}, \citealp{Bell2018Cosmic-rayEstimates} and many more). See \citet{Ferrari2009}, \citet{Blasi2013TheRays}, \citet{Deligny2016} and \citet{Amato2017CosmicReview} for recent reviews of the study of cosmic rays.\\

Cosmic rays are of great scientific interest for a variety of reasons. In general, (secondary) cosmic rays make up around 10\% of the background radiation we experience on earth (\citealp{UnitedNationsScientificCommitteeontheEffectsofAtomicRadiation2008SOURCESB}). High energy rays have long been a probe of fundamental physics, from the famous example of the effect of time dilation on muon decay rate (\citealp{Rossi1941VariationMomentum}, \citealp{Frisch1963Measurement-Mesons}), to more recently the new physics suggested by a lack of anisotropy in the cosmic ray electron-positron ratio - potentially from dark matter annihilation (\citealp{Aguilar2013FirstGeV}). High-energy cosmic rays can give valuable information about high-energy astrophysics, e.g. shockwaves in supernovae (\citealp{Giuliani2011}, \citealp{Ackermann2013DetectionRemnants.}), merging neutron stars (\citealp{Komiya2017iR/iMergers}) and potentially active galactic nuclei  (\citealp{ThePierreAugerCollaboration2007CorrelationObjects}, \citealp{PierreAugerCollaboration2017ObservationEV.}). Cosmic rays can be viewed as complementary messengers in multi-messenger astronomy, alongside photons, neutrinos and gravitational waves (\citealp{Branchesi2016Multi-messengerRays}). See \citet{Ginzburg1996CosmicReview}, \citet{Kotera2011TheRays} and \citet{Castellina2013AstrophysicsRays}, for recent reviews of the role of cosmic rays in astrophysics. More locally cosmic rays are a probe of solar physics (e.g. see \citealp{Potgieter2013SolarRays}), and are a key component of `space weather' (e.g. see \citealp{Turner2014}). Space weather can have an impact both on ground based communications systems (e.g. a modern day Carrington event, see \citealp{Love2016}) satellites and spacecraft electronics (\citealp{Choi2011AnalysisRelationships}) and organisms (including humans). Understanding the radiation environment is vital for understanding the impact of the dose astronauts receive on the International Space Station (ISS, e.g. the link between received dose and susceptibility to cataracts, \citealp{Cucinotta2001SpaceAstronauts.}, see \citealp{Cucinotta2007SpaceMissions} for an overview), or on a hypothetical voyage to Mars (\citealp{Kerr2013PlanetaryRiskier.}, \citealp{Zeitlin2013MeasurementsLaboratory.}).\\ 

The development of novel energetic particle detector technologies at the European Organization for Nuclear Research (CERN) provides an opportunity to improve measurements of cosmic rays both in space and on the ground. The turn of the 21st Century saw the advent of photon counting pixel detectors for radiation detection with the development of Medipix detectors (\citealp{Bisogni1998PerformanceChiplt/titlegt}, \citealp{Campbell1998ACounting}). Medipix detectors (now entering their fourth generation, Medipix1 \citealp{Amendolia1999MEDIPIX:Radiology}, Medipix2 \citealp{Llopart2002Medipix2:Mode}, Medipix3 \citealp{Ballabriga2011CharacterizationChip}, Medipix4 collaboration founded 2016) can detect and differentiate between many types of ionizing radiation, and present many advantages compared to other methods - but at the cost of a small collecting area and comparatively high expense. Medipix detectors have been used in a wide range of applications, including high-energy physics experiments (\citealp{Greiffenberg2009DetectionNeutrons}, \citealp{Vykydal2009TheDetector}, \citealp{Collins2011TheUpgrade}), medical physics (hence `Medi'€, e.g. \citealp{Blanchot2006Dear-Mama:Applications}, \citealp{Butzer2008MedipixPCA}, \citealp{Martisikova2011MeasurementTimepix}, \citealp{Jakubek2011ImagingDetector}, \citealp{Hartmann2013AHeiDOK}) and small animal imaging (\citealp{Accorsi2008High-ResolutionDetector}).\\

More recently, there has been increased interest in their application in space (\citealp{Kroupa2015ADosimetry}, \citealp{Granja2016TheOrbit}, \citealp{Gaza2017ComparisonEFT-1}, \citealp{Urban2017VZLUSAT-1:Application}). In particular their ability to distinguish between different particle types and give angular information have, for example, proved valuable in understanding the radiation environment of the ISS. Using Medipix in space was first discussed in \citet{Pinsky2011}. Seven NASA/IEAP-developed Radiation Environment Monitors (REMs), Timepix detectors in compact USB mounting, have been deployed to the ISS (altitude $\sim$400km), see \citet{Turecek2011SmallStation}, \citet{Kroupa2015ADosimetry} and \citet{Stoffle2015Timepix-basedStation}. Four of these have been in near continuous operation since 2012, operated via an onboard laptop. The second use of Medipix in space was on the European Space Agency (ESA) Proba V mission, launched on the 7th May 2013 to Low Earth Orbit (LEO) with an altitude of 820km, with the spacecraft payload Space Application of Timepix Radiation Monitor (SATRAM, \citealp{Granja2016TheOrbit}) onboard. SATRAM carries a single Timepix detector and is operating and continuously taking data today. In addition, Exploration Flight Test 1 (EFT-1), the first flight of the Orion Multi-Purpose Crew Vehicle (MPCV) on a two orbit, 4.5 hour trip on the 5th December 2014 took Medipix data the farthest from Earth to date at $\sim$5910 km (\citealp{Gaza2017ComparisonEFT-1}). Most recently, on the 23rd June 2017 the cubesat VZLUSAT-1 (altitude 510km, \citealp{Daniel2016TerrestrialCubeSat}, \citealp{Urban2017VZLUSAT-1:Application}) carried a miniaturised x-ray telescope, that uses Timepix detectors (\citealp{Baca2016MiniaturizedDetector}), into orbit for astrophysical, space weather studies, and terrestrial X-ray monitoring applications (see \citealp{Pina2015X-rayCubesat}).\\

In this paper we report the first results from one of the early uses of Medipix in orbit (and the first on a commercial platform, and the first with Medipix detectors in a 3D configuration), the Langton Ultimate Cosmic ray Intensity Detector (LUCID) on board TechDemoSat-1. The dominant source of particles detected by LUCID are trapped electrons and protons, and the instrument is designed such that it could be deployed as a hosted payload for satellite environmental monitoring. We discuss the design of the instrument and its operations, a new platform TAPAS for managing the large amounts of data produced by the experiment and a new machine learning algorithm for the automated classification of particles which can be used for other Medipix applications. The first results and some early science applications (e.g. mapping out the South Atlantic Anomaly) are also presented. In addition, LUCID is linked to a extensive programme of education and research in the classroom, CERN@school, where students can use Timepix detectors for both novel tests of traditional classroom experiments (e.g. inverse square law, see \citealp{Whyntie2013InvestigatingExperiment}) as well as original science, for example the Radiation In Soil Experiment (RISE) which has measured the radiation in different geological samples across the UK, and construction of a robotic three-dimensional radiation scanner (\citealp{Whyntie2016CERNschool:Classroom}, see also \citealp{Colthurst2015ResearchPhysics}, \citealp{Parker2017RealClassrooms}, \citealp{Parker2018IRISCommunities} for further general information). All of these projects are managed through the same data storage and reduction pipeline used in this work, see section \ref{sec:TAPAS}. Cubesats have already been shown to be highly effective educational tools e.g. \citet{Li2013}, and other Medipix devices have been used in the CERN@school programme e.g. RasPIX \added{\cite{Raspix}.}\\

In addition to LUCID, the ISS REMS, SATRAM and the VZLUSAT-1 x-ray telescope, there are also future planned deployments, see \citet{Pinsky2016An4}. Future missions include; a particle telescope architecture containing two Timepix detectors in sync, on the Rapid International Scientific Experiment Satellite (RISESAT, \citealp{Kuwahara2011SatelliteAvionics}, \citealp{GRANJA2014Timepix-BasedRISESAT}), a Japanese FIRST mission to orbit at $\sim$700km, further Medipix being sent to the ISS, HERA monitors (units containing single Timepix detectors being developed at NASA for use on future MPCV missions), the proposed Biosentinel astrobiology deep-space cubesat mission and on trans-lunar NASA-ORION missions in the 2020s.

The paper is organised as follows. In section \ref{sec:payload} we describe the LUCID payload. In Section \ref{sec:data_analysis} we describe the data reduction process, using GridPP resources, processing all the data that was produced by the payload with a new machine learning analysis framework. We also describe the development of the Timepix Analysis Platform at School (TAPAS), a web platform to present the reduced LUCID data to secondary school students. In Section \ref{sec:results} we give the preliminary results that we have gathered based on this data. In Section \ref{sec:discussion} we discuss our results and  future planned work. Finally, in Section \ref{sec:conclusions}, we summarise our findings.\\

\section{The LUCID Payload} \label{sec:payload}
\subsection{LUCID and TechDemoSat-1}

\begin{figure*}
\includegraphics[scale=0.7]{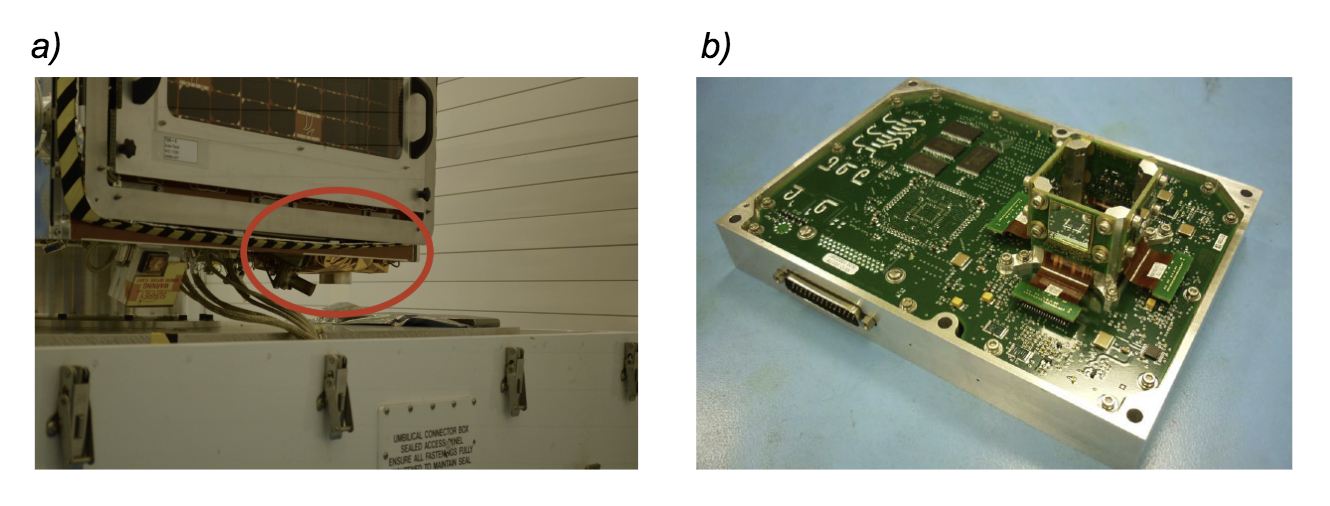}
\caption{LUCID shown integrated on TDS-1 in the SSTL assembly room b) the LUCID payload showing the Timepix detector array (taken from the LUCID System Design Document). The four orthogonally positioned detectors are visible on the right hand side of the instrument.}
\label{fig:lucid_image}
\end{figure*}

LUCID is a payload on the technology demonstration satellite TechDemoSat-1 (TDS-1, see figure \ref{fig:lucid_image}). The project started in 2008, and was developed as a collaboration between Langton Star Centre secondary school student researchers, the Medipix Collaboration, and Surrey Satellite Technology Limited (SSTL), who built both LUCID and TDS-1. LUCID is part of the TDS-1 Space Environment Suite, which consists of the Miniature Radiation Environment and effects Monitor (MuREM, \citealp{Taylor2012TheTechDemoSat-1}, \citealp{Underwood2016DevelopmentPayloads}), the Charged Particle Spectrometer (ChaPS, \citealp{Kataria2013}) and the Highly Miniaturized Radiation Monitor (HMRM, \citealp{Mitchell2014TheMonitor}, \citealp{Guerrini2013DesignHMRM}). TDS-1 launched on 8 July 2014 (15:58:28 UTC) on a Soyuz-2-1b launch vehicle with Fregat-M upper stage from the Baikonur Cosmodrome in Kazakhstan, into a 635 km, 98.4$^\circ$ Sun-synchronous orbit. LUCID began data collection shortly after launch, and data collection ceased on the 4th July 2017. TDS-1 operations have now ended, and at some point in the medium-term it will be deorbited by the Icarus-1 Cranfield Drag Augmentation System de-orbiter (\citealp{Hobbs2013De-OrbitTechDemoSat-1}) which will over the next 25 years guide the spacecraft into the Earth's atmosphere, where it will disintegrate.\\

The detectors used in LUCID are based on the Timepix ASIC chip (\citealp{Llopart2007TimepixMeasurements}, \citealp{Plackett2010}), part of the second generation of Medipix (Medipix2, \citealp{Llopart2002Medipix2:Mode}, \citealp{Llopart2007TimepixMeasurements}; \citealp{Ballabriga2011CharacterizationChip}). The detectors used are equipped with a 300$\mu$m silicon sensor. The Timepix chips contain $256 \times 256$ pixels, each measuring 55$\mu$m on each side, giving a total collecting area of 1.98cm$^2$.  The circuitry to digitise the output of each pixel is contained within the footprint of the pixel, meaning only digital information is transferred out of the pixels. Timepix can operate in event counting mode (the base functionality of Medipix2, where a counter is incremented each time that during the shutter time period the charge deposited is over the designated threshold), arrival time mode (where essentially the first time during the shutter time period that the charge over a threshold is deposited is recorded) and energy sensitive Time over Threshold (ToT) mode (measuring the signal amplitude providing the per-pixel deposited energy). LUCID was always run in ToT mode.\\

\subsection{LUCID Instrument Design}
We briefly summarise the instrument design here; a full technical overview and detailed design information of LUCID is in the LUCID System Design Document (D. Cooke, SSTL, private communication). The payload has five Timepix radiation detectors in a cube-like configuration (see Fig. 1b), with four detectors orthogonally positioned facing outwards (TPX0 through TPX3), and the fifth in the centre (TPX4), facing outwards (relative to the centre of LUCID). A photograph of the instrument is shown in Figure \ref{fig:lucid_image}. The chips were surrounded by a 0.75mm thick aluminium dome which blocks intense light, plasma and low-energy charged particles. LUCID is mounted on the `Earthside'€ of TDS-1.\\

The detectors were calibrated by the Institute of Experimental and Applied Physics (IEAP) at the Czech Technical University in Prague. The calibration process involves exposing the detectors to 'X rays of discrete energy, and modelling the low energy end non-linear response of each individual pixel, see \citet{Jakubek2008PixelParticles} and \citet{Jakubek2011PreciseMode}.\\

The performance and expected measurements of LUCID were simulated in \citet{Whyntie2014SimulationEnvironment}, \citet{Whyntie2015FullEnvironment}, who simulated the environment of LUCID and expected data rates using European Space Agency'€™s (ESA) SPace ENVironment Information System (SPENVIS, \citealp{Heynderickx2004NewSPENVIS}) and GEANT4 (\citealp{Agostinelli2003Geant4aToolkit}, \citealp{Allison2006Geant4Applications} and \citealp{Allison2016RecentGeant4}).

\begin{table*}
	\caption{Payload Specifications}
	\label{table:specs}
	\begin{tabular}{|p{0.2\linewidth}|p{0.75\linewidth}|}
	Detectors & 5 Timepix hybrid pixel detectors \added{(300$\mu m$ silicon)}, 1 RADFET \added{(metal-oxide-silicon)}\\
	Power & 1 permanent 5V supply and 1 28V supply for the Timepix detectors. With all 5 detectors running, maximum power usage is 8W\\
	Storage & 2GB NAND flash\\
	Data transfer & Maximum 20Mbit/s via X-Band\\
	Physical Dimensions	& 220mm x 135mm x 33mm (SSTL nano tray)\\
	Mass & 1.2kg\\
	Dome & 0.75mm aluminium, energies excluded; $E_e > 0.4MeV$ and $E_p > 10.0MeV$ \citep{Whyntie2015FullEnvironment}\\ 
	\end{tabular}
\end{table*}

\subsection{LUCID Operations}

\begin{table}
	\caption{Timepix configuration DAC-parameters, \added{used for a large number of runs on all detectors} (ID 321)}
	\label{table:config}
	\begin{tabular}{|p{0.5\linewidth}|p{0.4\linewidth}|}
		Active detectors & 0, 1, 2, 3, 4 \\
		Frame rate & 1 per second\\
		Shutter Exposure Time & 0.3 s\\
		\added{Clock} & \added{33 MHz}\\ 
		Number of frames & 100\\
		Bias Voltage & 20.02 V\\
		IKrum & 1\\
		Disc & 127\\
		Preamp & 255\\
		DAC Code & 6\\
		Sense DAC & 0\\
		Ext. DAC Sel. & 0\\
		BuffAnalogA & 127\\
		BuffAnalogB & 127\\
		Hist & 0\\
		THL Fine & 300\\
		THL Coarse & 7\\
		VCAS & 130\\
		FBK & 128\\
		GND & 80\\
		CTPR & 0\\
		THS & 100\\
		BiasLVDS & 128\\
		RefLVDS & 128\\
	\end{tabular}
\end{table}

Although the payload has five detectors, for the majority of its time in orbit, only three were taking data at any one time. This was because LUCID would be drawing too much power with all five detectors taking data. Towards the end of operations all five detectors were switched on and taking readings as the other payloads entered different phases of operation.  The detectors are built and configured to operate in sync - i.e. they capture frames simultaneously. Specifications are shown in table \ref{table:specs}.\\

Each detector is able to independently capture a `€˜frame' of the radiation that passes through the detector over some time period. LUCID was run with a range of frame rates and shutter exposure times, with the most commonly used shown in table \ref{table:config}, and a frame taken every second. The payload has a 2GB NAND flash storage which is used to store data between passes of the ground stations, allowing LUCID to take data for more extended periods of time. The amount of available storage space and data transfer requirements of other experiments dictated frame rate.\\

LUCID operated in a data gathering capacity from late October 2014 until July 2017. However, the first few months of operation were dedicated to instrument commissioning, and were used to find the optimal configuration settings for data recording and the abilities of the payload. Nominal operations commenced in April 2015. During this time, over 2.1 million frames were captured, over 82 runs and 11,700 files. A run is an 8 day capture period, defined by the operational schedule of TDS-1.\\

The final data product from LUCID consists of 256x256 PNG images of each frame, CSV files that include the particle counts on a per frame basis, x,y,C formatted files for each frame, together with associated metadata and navigation and time stamp. This is post-conversion, as data was retrieved in a proprietary compressed raw format.

\section{LUCID Data Analysis} \label{sec:data_analysis}

Different particles typically produce very different tracks when they pass through a Medipix detector (e.g. \citealp{Llopart2002Medipix2:Mode} and \citealp{Bouchami2011MeasurementDevice}). Given the large volume of data collected by the instrument, it is necessary to automate the detection of different particles. In this section we present two new high-performing machine learning algorithms for classifying tracks in LUCID frames into particle types. This is a difficult task both because of the large quantity of data (meaning that the algorithm has to be reasonably fast, and the task cannot be done solely by human classifiers), as well as the general difficulty in classifying tracks (essentially a pattern recognition problem e.g. \citealp{Holy2008}).\\

Firstly a clustering algorithm is run on the LUCID frames to identify individual particle tracks.  After having identified the track for each individual particle, we considered two main methods classifying particles:

\begin{itemize}
\item `Metric-Based Network' (MBN): we extract a given set of features for each particle track and then classify the particle based on these metrics using machine learning. This network does not take per-pixel energy distribution into account.
\item `Deep learning'€: to use convolutional neural networks to classify the particles tracks directly with a pure deep learning classification algorithm. This network does take per-pixel energy distribution into account.
\end{itemize}

Both of these methods are supervised machine learning algorithms and therefore the main limiting factor for the development of the classification algorithms is the computing power available and the amount of labelled data that had been collected. In this work the training data is labelled by human classifiers.\\

We also compare the performance of our algorithms to an analytic classifier (i.e. tracks are classified based on an analytic function of the feature metrics) used in an early stage of the analysis of the data (see \citealp{Whyntie2015CERNschool:Detector}).\\

Our approach can be compared to other studies of pattern recognition and cluster analysis in Timepix detectors, such as \citet{Vilalta2011DevelopmentDevices}, \citet{Hoang2012LETDetector}, \citet{Opalka2013LinearTarget} and \citet{Holy2008}. Other possible machine learning approaches to classifying particles not considered here include generative adversarial networks as well as autoencoders.\\

\subsection{Technical Implementation}

The payload was operated by submitting Payload Task Request (PTR) files to SSTL. These files contain information about the start time of the run, the configuration file to use, and the overall schedule for capturing data in this run. The LUCID configuration files specify which detectors should be used in this capture and the settings of each detector to be used, for instance the shutter exposure time and threshold value.\\

During the operating lifetime of the payload, data was transferred from the satellite to SSTL and then downloaded to an IRIS server at the Langton Star Centre in Canterbury, UK. A cron job (a system used for scheduling jobs on UNIX based systems) ran every day to check for the existence of new files on the SSTL FTP server, and if files were detected, they were downloaded and processed immediately, converting them from a proprietary format for LUCID into individual frames of x, y, C files (x-coordinate, y-coordinate, \deleted{charge} \added{ToT value}; tab separated). Metadata (time of capture, detector frame was from, the file the frame belongs to and location information) was stored in an SQLite database.\\

Pre-processed data was compressed and uploaded to GridPP (\citealp{Collaboration2006GridPP:Physics}, \citealp{Britton2009GridPP:Physics})  storage, using the CernVM (\citealp{Buncic2010CernVMApplications}). At the same time as the uploading process, custom software was developed to process runs in parallel, allowing many runs to be analysed at the same time, exploiting the highly parallel nature of running jobs on GridPP.\\ 
 
The processing software has been developed in Python, using Tensorflow (\citealp{Abadi2016TensorFlow:Systems}) machine learning for particle analysis (see sections \ref{sec:MBN} and \ref{sec:deep}).  The CernVM File System (CVMFS) was used to deploy the software dependencies and Python interpreter to the grid worker nodes to run the software, and jobs were submitted using the gLite middleware to a specific GPU-backed job queue. The full source for the software is available at: \url{https://github.com/willfurnell/lucid-grid/}.\\

Each job running on a worker node copied data from a storage element to its working directory, extracted it and then ran the classification algorithm. This process included getting frames and their metadata (namely the capture time) from the pre-process database, and then calculating which of the Two Line Element files (TLE, files that contain the information to track the location of the satellite\footnote{`A NORAD two-line element set consists of two 69-character lines of data which can be used together with NORAD's SGP4/SDP4 orbital model to determine the position and velocity of the associated satellite.' - \url{https://celestrak.com/columns/v04n03/}}) had a date closest to the frame'€™s timestamp. This allowed the latitude and longitude of LUCID when the capture took place to be calculated. The classification was then run on all frames in the particular file, and for each frame alpha, beta, X-ray, proton, muon and other particle classifications, and the latitude, longitude, frame number, capture timestamp were submitted directly to the TAPAS (section \ref{sec:TAPAS}) database via a POST request to a REST API endpoint. Each run and file (with its timestamp, ID and configuration file used) was also submitted this way. 

\subsection{The TAPAS Data Analysis and Visualisation Tool} \label{sec:TAPAS}

We have developed The Timepix Analysis Platform at School (TAPAS) to allow secondary school student researchers across the UK to analyse and share the data that they gather using Timepix radiation detectors across all CERN@school projects, and additionally as a home for the particle count data from the LUCID experiment. The platform allows users to upload their own data, taken with the Timepix detectors using a software package called Pixelman (\citealp{Turecek2011Pixelman:Detectors}), or data which has been provided to them, such as the TimPix ISS-REM radiation data. A web application was chosen instead of a desktop application because it  is very easy to access - students only need a web browser and internet connection to use the service.\\

Once a user has uploaded a dataset, a cron job is used to run the analysis service every 10 minutes to process the data. The analysis service uses multiple processes to analyse large amounts of the data in a parallel fashion, using the lucid\_utils LUCID algorithm. The service also generates an image of every frame processed. TAPAS has also been used to analyse small amounts of the LUCID data (the whole LUCID data set required the resources of GridPP).\\

TAPAS is a Django framework based web application, primarily written in Python for the page generation and HTML, CSS and JavaScript for the frontend user facing elements. The web platform has a MariaDB database backend that is used to store all metadata relating to a LUCID run, file and frame and particle classifications. This database also includes information and analysis results relating to user-made uploads.\\

The platform includes an API, using the `Django Rest Framework'€™ to allow uploads and particle classifications to be submitted using a programmatic REST (Representational state transfer) interface. This is of particular importance to the LUCID data, as this is how particle classifications were submitted to the database from the software running on GridPP worker nodes.\\

We have programmed TAPAS to allow students to download CSV files with particle count data, allowing them to conduct further analysis using their own choice of software packages - or even by choosing to write their own, in a programming language they are most comfortable with using. All of the LUCID particle count data is downloadable as CSV files from TAPAS.\\

\subsection{Event Identification} \label{sec:identification}

\begin{figure}
\centering
\includegraphics[scale=0.6]{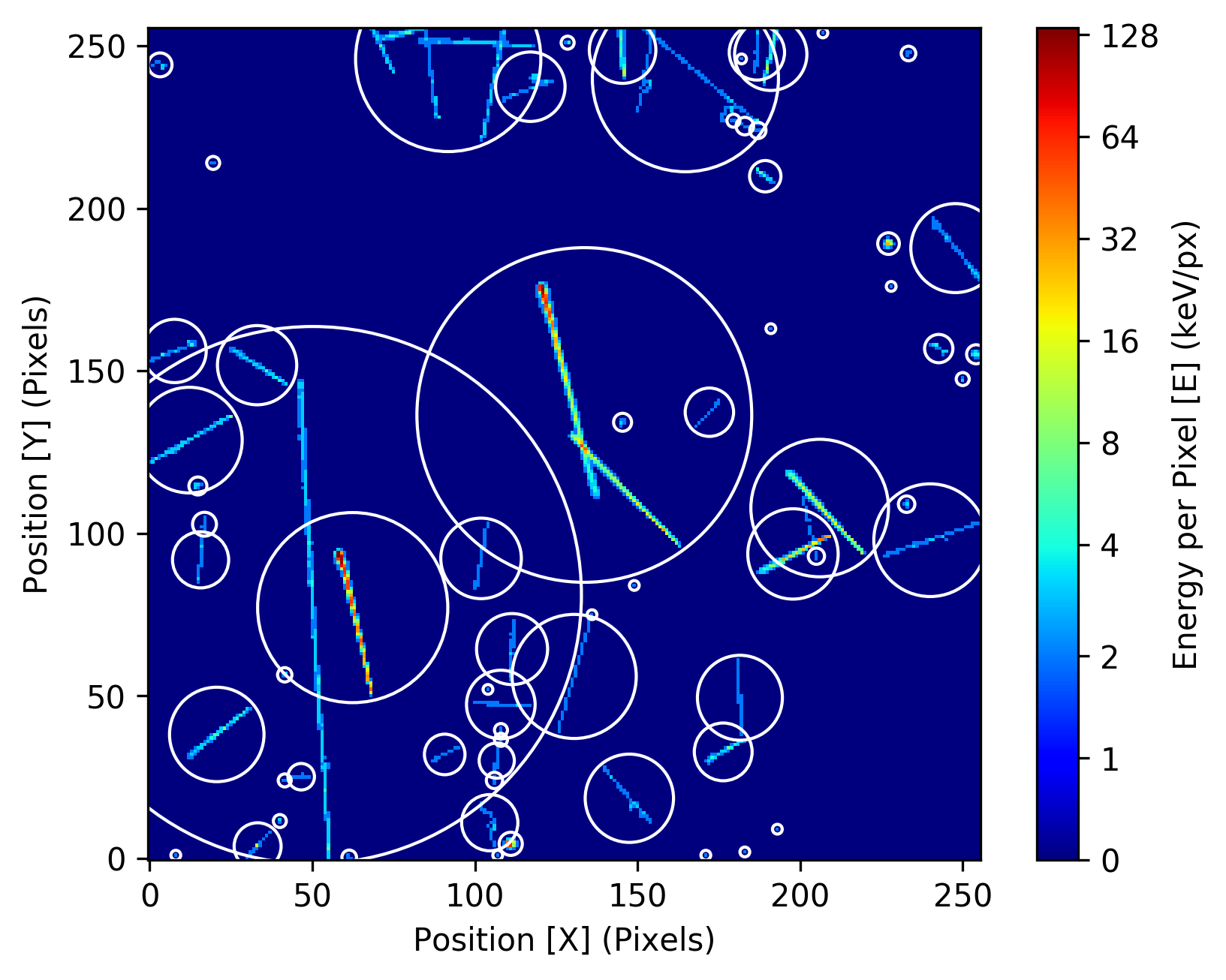}
\caption{An example LUCID frame with individual particle tracks identified.}
\label{fig:sample_tracks}
\end{figure}

The first step of classifying particles in a given Timepix detector frame is to identify each individual particle track. The frame is fed in as a 256 x 256 matrix of energy values. Non-zero pixels are systematically selected, and any non-zero pixels in the `ring'€™ of the 8 adjacent (including diagonals) pixels are then added to the associated cluster. For each adjacent pixel, all other adjacent pixels are checked until there are no more pixels with non-zero energy values. The clustering algorithm works through the entire frame and returns a list of clusters, each of which is a list of pixels. \added{An alternative clustering method is discussed in \citet{Granja2018}, which uses track shape and deposited energy values to classify the particle type, energy, and direction of events.} Figure \ref{fig:sample_tracks} shows an example LUCID frame where each particle track has been successfully identified using the clustering algorithm.\\

\subsection{Training Data}

\begin{figure*}
\centering
\includegraphics[scale=0.6]{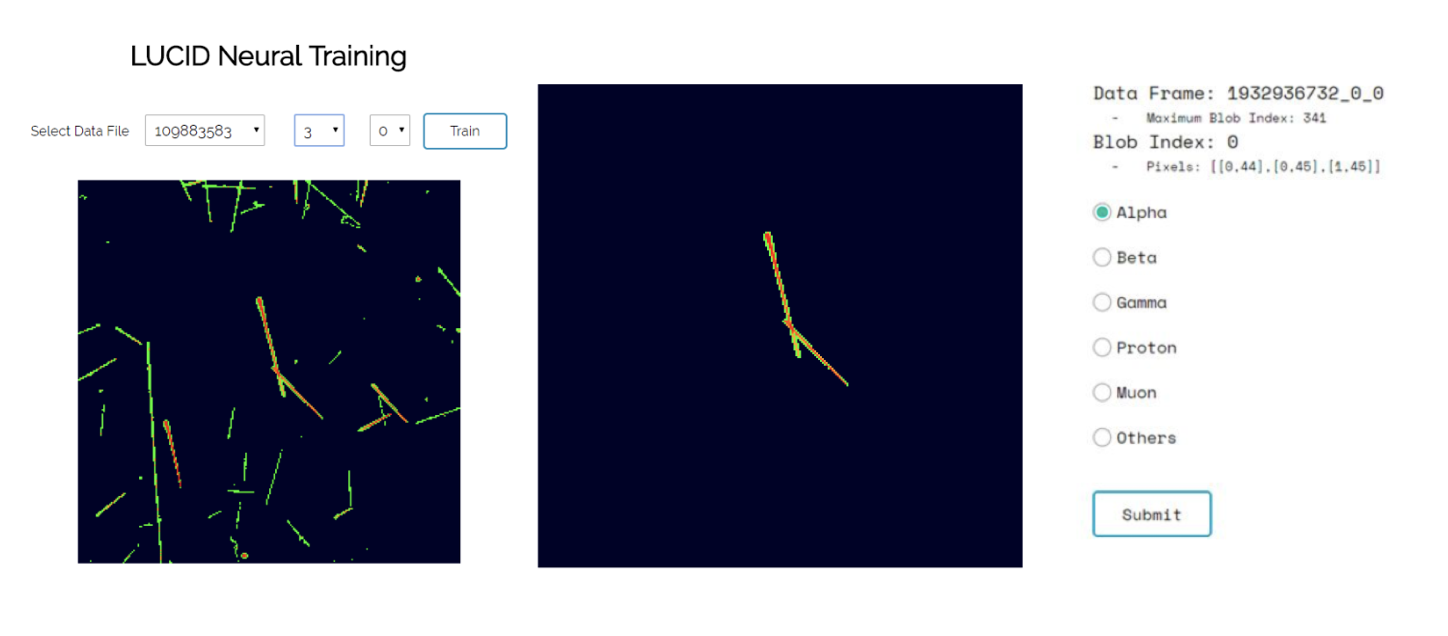}
\caption{Screenshot from volunteer classification questionnaire, as seen by the student researchers.}
\label{fig:questionnaire}
\end{figure*}

Both algorithms considered here are supervised, and require that a subset of the data is labelled with the `true' classification. To generate training data for classifying particle tracks, a web application called LUCID Trainer was created (Figure \ref{fig:questionnaire}). It allows volunteer classifiers (typically IRIS secondary school researchers) to simply click through automatically generated questionnaires. The responses would be stored in a database alongside the pixel data and the metadata for finding out which frame the cluster belonged to.\\

The LUCID Trainer web application is written in PHP for dynamic page generation and HTML, CSS and JavaScript for the frontend user interface. All the LUCID data is accessed via a custom REST API connecting the LUCID metadata database to the web application. The user response data is sent via an AJAX request to another PHP file that then stores the response in a simple SQLite database which was used due to its lightweight requirements.\\

The training set used consisted of 1800 particles (tracks identified as in section \ref{sec:identification}). These tracks were each classified once by a student researcher, with 24 classified as alpha particles, 988 as beta/electrons, 547 as X-ray/photons, 27 as muons, 160 as protons and 54 as `€˜other'€™. Clusters that the user could not identify, or overlapping clusters were those classified as `other'. The student researcher classifiers themselves were trained using example tracks with known particle type, and reference diagrams similar to Figure 1 in \citet{Bouchami2011MeasurementDevice} and Figure 1 in \citet{Whyntie2015CERNschool:Detector}. The student researchers, and therefore the algorithm, may have had issues separating \deleted{high energy photons from electrons.}\added{one particle type from another.} 
The resulting training set is thus based on human classifications, which as a methodology is necessarily less accurate than using a training set constructed from known calibrated sources. \deleted{Therefore quoted particle counts and fluxes for LUCID calculated in this way necessarily have some caveats.} Nevertheless, these preliminary initial results still give us a good overview of the distribution of morphology of detections, and in the future calibrated classifications can be generated and used for training to obtain more realistic particle counts.\\

The nature of supervised machine learning algorithms trained on human classifications means that the algorithm can at best reproduce the classification that a human classifier would give. Similar projects such as Galaxy Zoo (e.g. \citealp{Lintott2008GalaxySurvey}, \citealp{Willett2017GalaxyImaging}, \citealp{Smethurst2016GalaxyGalaxies}) and the rest of the Zooniverse have shown success in the classification of large samples of image and image-like data based on a training set of human classifications. \citet{Vilalta2011DevelopmentDevices}, in contrast to this work, trained their algorithm using data taken at the Heavy Ion Medical ACcelerator Facility (HIMAC) in Chiba, Japan e.g. they fired beams of known particle type and energy at Medipix detectors, so they had a set of tracks labeled by the true particle properties (see also \citealp{Jakubek2010DetectionTimepix}). Our human labelling of tracks will be imperfect - however human labelling is still a valuable approach to develop for the classification of tracks as human classifiers can identify `unexpected' tracks that might be in the LUCID data but not in a laboratory produced training set (e.g. \citealp{Beck2018IntegratingTasks} found that a combination of human classification and machine learning classification gave the best results).\\

\subsection{Metric Based Network} \label{sec:MBN}

The Metric Based Network (MBN) approach to classifying the particles was to calculate a small number of easily computed features for each track, and to then classify the particles based on the metrics that had been calculated.\\

The primary classifier used was the multi-layer neural network. The hyperparameters (number of hidden layers, number of nodes in each hidden layer etc.) for this architecture was optimised manually. Other machine learning classifiers were used for comparison as well as for ensuring that the limiting factor for the accuracy was the size of the training set and not the architecture.\\

The general principles behind the multi-layer neural network are described in \citet{Haykin1998NeuralFoundation}, and see the extensive literature on similar problem of classifying handwritten digits (a similar problem) e.g. \citet{McDonnell2015FastAlgorithm}. See \citet{Denby1988NeuralPhysics} and \citet{Peterson1989TrackNetworks} for early uses of neural networks in studying particle tracks in high energy density physics experiments, and \citet{Farrell2017TheTracking} for a more recent example in the LHC.\\

The eight metrics calculated from the pixel cluster representing the particle track for the algorithm are:

\begin{itemize}
\item Number of Pixels -†' This is calculated as the length of the pixels list.
\item Radius - This is calculated using the calculate radius function.
\item Density - This is calculated using the calculate density function.
\item Line Residual -†' This is calculated using the calculate non-linearity function.
\item Curvature Radius and Circle Residual - These are calculated using the find best fit circle function.
\item Average Neighbours - This is calculated using the find average neighbours function.
\item Width - Number of Pixels / Diameter (\textit{if} number of pixels $>1$ \textit{else} width = 0)
\end{itemize}

The precise definitions of these features are given in Appendix B. In early analyses of the data these metrics were used for an analytic classification of the tracks (e.g. if number of pixels less than 4, then classify as a X-ray etc.), also described in Appendix B.\\

\begin{figure}
\centering
\includegraphics[scale=0.5]{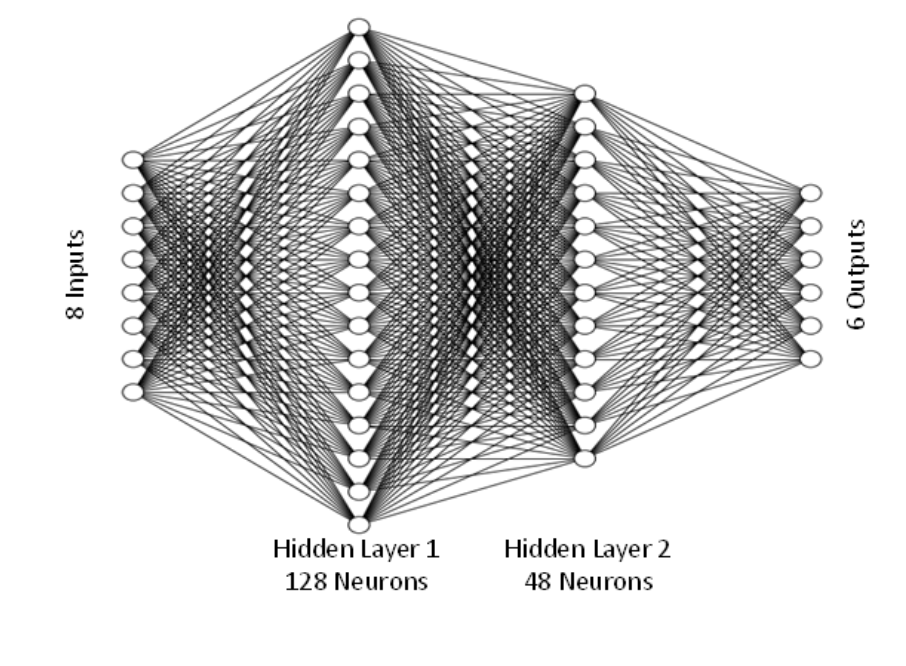}
\caption{The architecture of the SML neural network used (diagram not to scale)}
\label{fig:network_structure}
\end{figure}

The final multi-layer neural network architecture has 8 inputs in the input layer (for the 8 metrics),  128 nodes in the first hidden layer with the sigmoid activation function, 48 nodes in the second hidden layer with the sigmoid activation function, and 6 outputs in the output layer with the softmax activation function, see figure \ref{fig:network_structure}. No dropout or regularisation was needed as the size of the training dataset was already limited. The Gradient Descent Optimiser was used as the training algorithm. The network was trained with a batch size (the number of training examples in one forward/backward pass) of 128 particles. Additional classes can be added to the Metric-Based Network with ease as the outputs are one-hot encoded however it will have to be trained from scratch on the dataset.\\

\subsection{Deep Learning} \label{sec:deep}

Deep learning algorithms present the possibility to classify the particle tracks directly from the images (which contain more information) rather than from the pre-identified features considered in the MBN algorithm. Deep learning approaches have high requirements of computing power and typically are used with GPUs, so we only present some early results here, although the approach could lead to much higher levels of accuracy with future work.\\

The deep learning architecture considered was a Deep Convolutional Neural Network (CNN e.g. see \citealp{Rawat2017DeepReview} for a recent review) run for 5000 epochs. Epochs are one forward pass and one backward pass of all the training examples. We do not describe the algorithm in great depth as the method remains preliminary. A Deep Residual Network (DRN e.g. \citealp{He2016DeepRecognition}) has also been tested on the data, but as of yet only for a small number of epochs so has not yet attained a comparable high level of precision, but may be competitive in the future.\\

Other advantages of the Deep Convolutional Network to the Metric-Based Network is that the Deep Convolutional Network allows the energy values to be considered (the feature metrics use for the MBN algorithm don't use the energy value of the pixels, just whether the pixel was non-zero or not) and the convolutional filters allow humans to visualise what the network has learned.\\

\subsection{Algorithm Performance}

\begin{figure}
\centering
\includegraphics[scale=0.5]{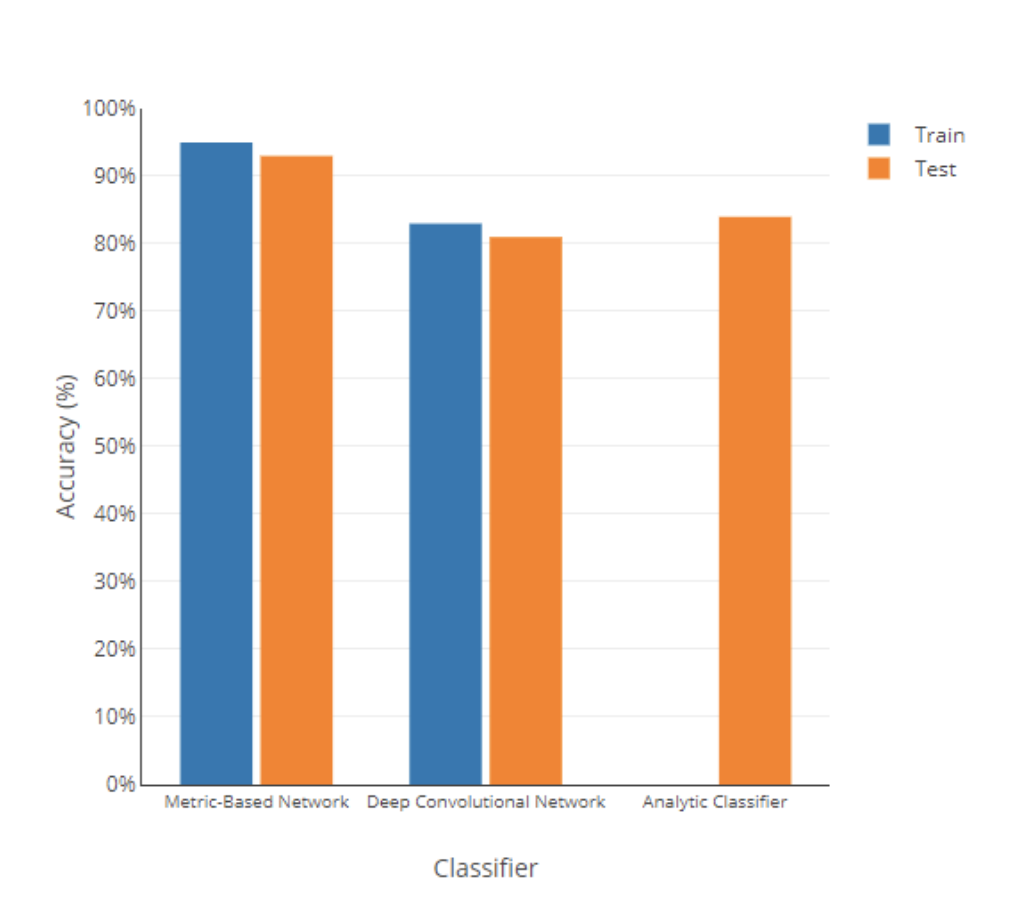}
\caption{Overall accuracy of each algorithm}
\label{fig:overall_accuracy}
\end{figure}

\begin{figure*}
\centering
\includegraphics[scale=0.5]{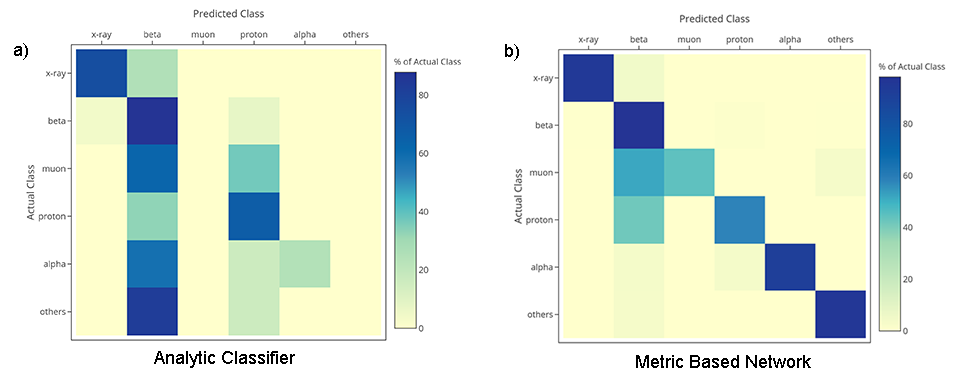}
\caption{Confusion matrices for the analytic classifier and MBN. Squares in the grid are colour coded by percentage of `€œactual class'€ classified as `predicted class'€.}
\label{fig:confusion_matricies}
\end{figure*}

Figures \ref{fig:overall_accuracy} and \ref{fig:confusion_matricies} show the performance of the algorithms, the Metric Based Network (MBN) Deep Convolutional Network (DCN) and the Analytic Classifier (AC). Other `€œoff the shelf'€ machine learning classifiers (support vector machines, k-nearest neighbours, a decision tree and random forest, all available in Tensorflow) were also tested to confirm that our network architecture was optimal but are not shown in the figure. The hyperparameters for these algorithms were not investigated and the default parameters were used.\\

Figure \ref{fig:overall_accuracy} shows the percentage correctly classified for a) the Metric-Based Network (MBN), b) a Deep Convolutional Network (DCN) and c) the Analytic Classifier (AC).\\

All performed between 80-95\% accuracy, with the MBN performing best on the test data. The AC had a high accuracy of 84\% but still less accurate than the MBN. This is because the dataset consists of much more beta and X-ray particles and therefore the overall accuracy appears to be high while its performance on a particle-by-particle basis was much poorer (see Fig \ref{fig:confusion_matricies}b). Figure \ref{fig:confusion_matricies} shows the confusion matrices for our best algorithm, the Metric-Based Network, and the Analytic Classifier to show how machine learning leads to much better performance. Each square shows the probability one type of particle has of being classified as another. Complete success in classification would correspond to a diagonal. The Metric-Based Network algorithm performs well, with the only substantial misclassifications being ~30-40\% of muons and protons being misclassified as electrons - partially because of similarities in the track shapes, partially because electrons dominate the overall sample, and partially because the labels used in the training are imperfect. The Analytic Algorithm performs much poorer, misclassifying all particles as either electrons and protons a substantial fraction of the time, and failing to identify any muons or `€œOthers'€.\\

\section{Results} \label{sec:results}

In this section we present some preliminary results from the data and particle classifications described in Section \ref{sec:data_analysis}. We use the classifications from the Metric-Based Network algorithm e.g. the particles are classified only based on track morphology, and some tracks will be in incorrectly classified, both due to the algorithm not having 100\% accuracy (see Fig \ref{fig:confusion_matricies}b) and imperfect labelling in the training set.

\subsection{Classifications}

Table \ref{table:classifications} presents the particle classifications that we have obtained for the whole lifetime of the LUCID payload. Different configuration files may produce different classifications due to saturating the frame (or conversely having completely empty frames), and therefore these results should be taken as a preliminary analysis, rather than confirmed results. Full payload data results are available on TAPAS \url{https://tapas.researchinschools.org}

\begin{table}
	\centering
	\caption{Event Classifications (all frames)}
	\label{table:classifications}
	\begin{tabular}{|p{0.2\linewidth}|r|}
		Alpha & 75902\\
		Beta & 169391026\\
		X-ray & 369816217\\
		Proton & 1065545\\
		Muon	& 569771\\
		Other & 6703109\\
	\end{tabular}
\end{table}

\subsection{Radiation Map and the South Atlantic Anomaly}

\begin{figure*}
\centering
\includegraphics[scale=0.4]{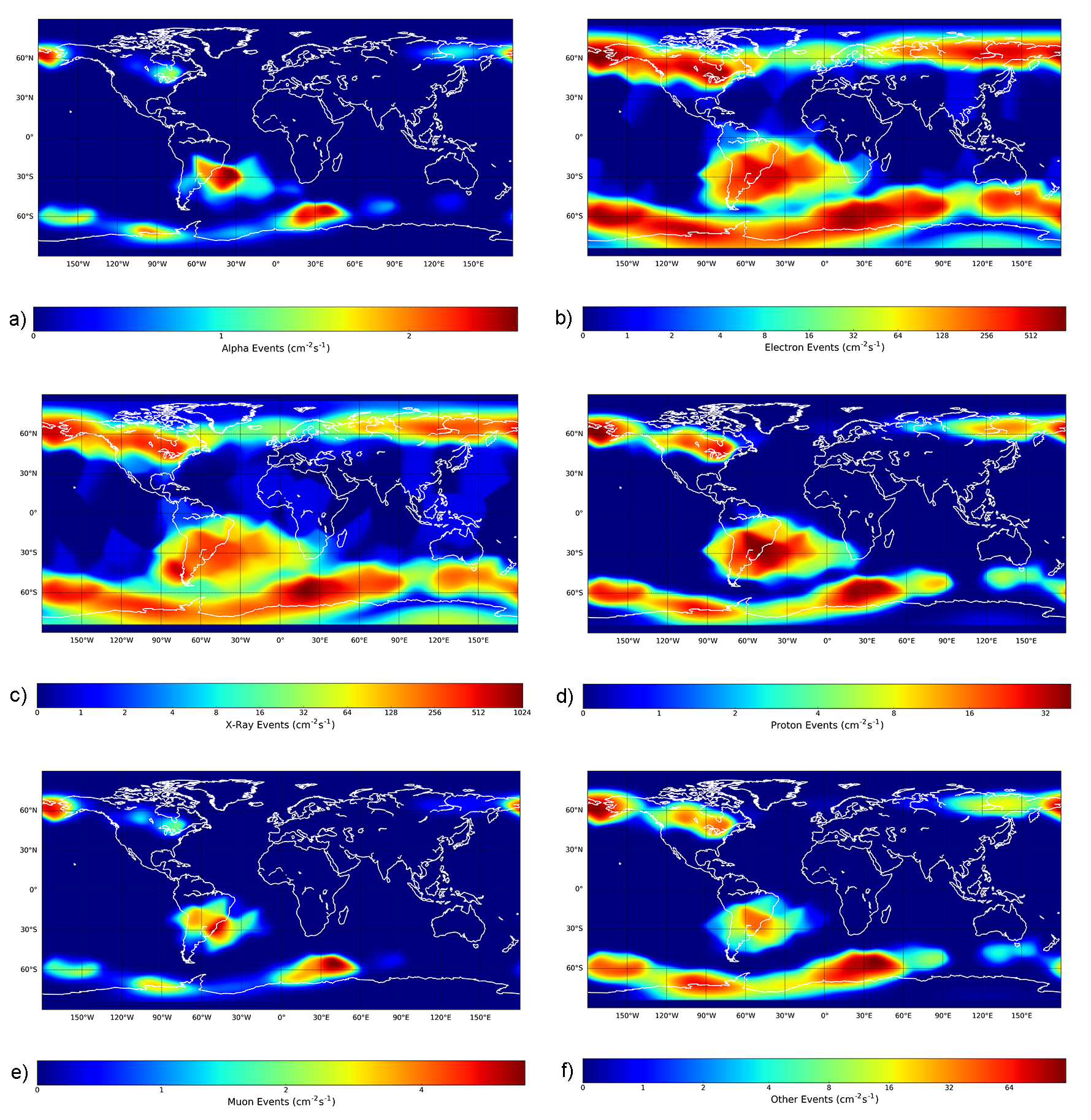}
\caption{Radiation maps for alpha particles, electrons, X-Rays, protons, muons and unclassified particles}
\label{fig:radiation_map}
\end{figure*}

\begin{figure*}
\centering
\includegraphics[scale=0.4]{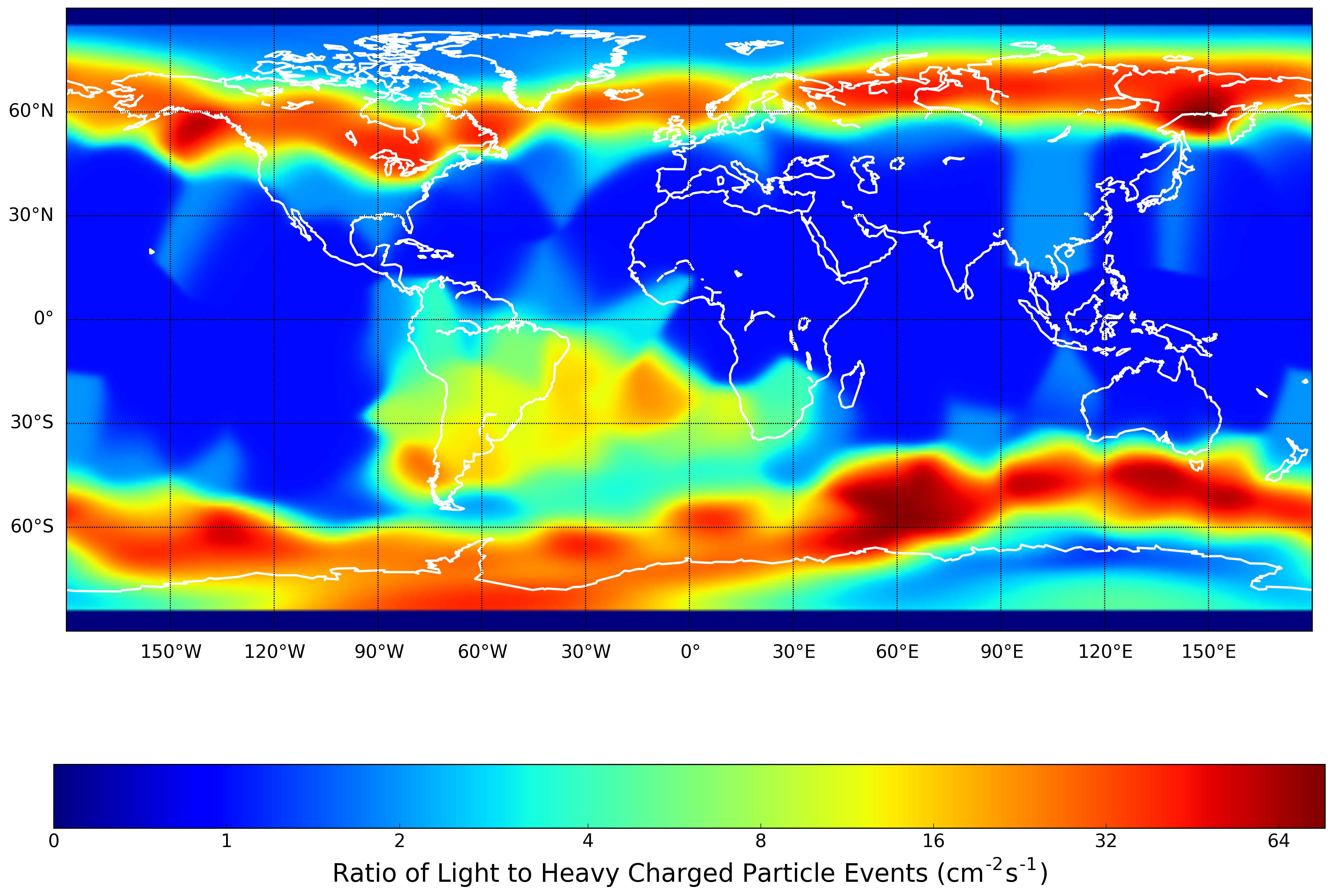}
\caption{A map of the ratio (Betas + Muons+1) / (Protons + Alphas+1). The +1 on the denominator to prevent division by 0 and +1 on numerator for a 1:1 ratio when both are 0.}
\label{fig:radiation_2}
\end{figure*}

We plot in figure \ref{fig:radiation_map} the number of particles detected over the Earth'€™s surface for $\sim$4000 frames for different particles. Higher radiation levels (by more than a factor of ten) around South Atlantic Anomaly (an area of known increased radiation flux, centred at roughly 30$^\circ$S, 60$^\circ$W) and the poles (roughly 60$^\circ$N, 60$^\circ$S) are clearly evident for all particle types. We also show in figure \ref{fig:radiation_2} the ratio of heavy charged to light charged particles - this is non-uniform, showing that that heavy charged particles have disproportionate intensity in the SAA and poles compared with the light charged particles.\\

The aim in this section is to illustrate that LUCID is giving sensible results, as opposed to give detailed measurements of dose, linear energy transfer (LET) and particle energy spectra.
Although LUCID can make estimates of the energies of particles because the detectors have been calibrated, this analysis is beyond the scope of this study and will be the focus of future work.\\

To create flux maps, a specific date range was chosen that covered the whole Earth, while also having a homogeneous set of configuration files (also limiting the effects of any time evolution of the radiation field). The sub-set of the data considered consisted of 404 files, from 2016-08-17 to 2016-09-21, where we took the first 10 frames from each of the files to result in a total of 4040 frames. For each file, the measured flux (count rate per unit time per unit collecting area) was calculated. Then a K-dimensional (K-D) tree (with a maximum number of neighbour lookups of 100) was used, so that for every given latitude and longitude reached by LUCID, the plotted flux at that point is the average of neighbouring frames flux, weighted inversely by the distance to the frames.\\

\subsection{Heavy Charged Particles}

\begin{figure*}
\includegraphics[scale=0.6]{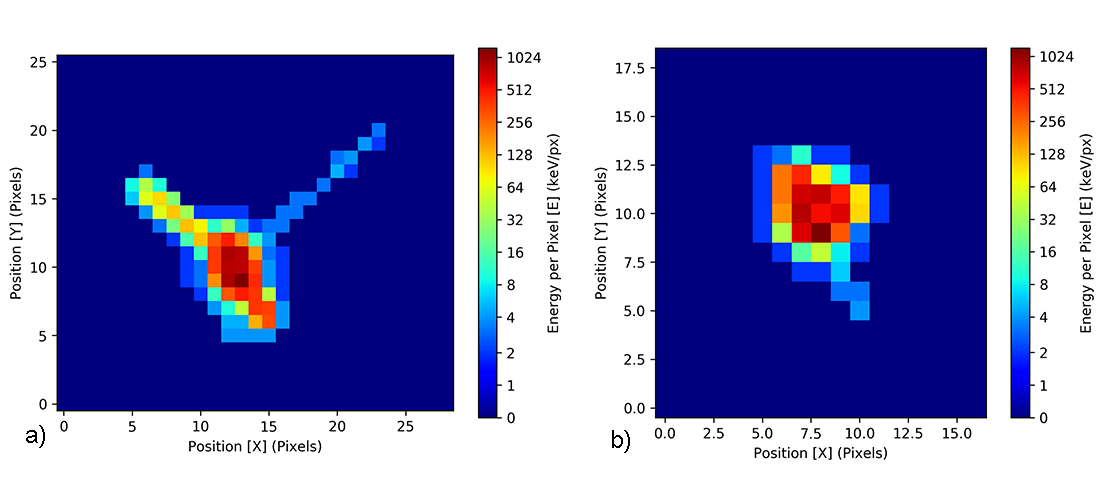}
\centering
\caption{Sample tracks of likely heavy charged particles. The colour scale is linear and scales with the energy deposited by the ion in the silicon layer for each pixel.}
\label{fig:heavy_ion}
\end{figure*}

Of our tracks, about 3\% were classified as `€œOther'€, and a substantial proportion of these particles are likely nuclei heavier than helium. We present example tracks in figures \ref{fig:heavy_ion}a and \ref{fig:heavy_ion}b, showing the extended shape. Timepix are particularly well suited for identifying heavy charged particles as they can extract track shape (as opposed to just particle energy), see \citet{Granja2011ResponseIons}, \citet{Stoffle2012InitialDetectors} and \citet{Hoang2012LETDetector}. Identifying flux levels of heavy ions carries important astrophysical information e.g. \citet{Aguilar2018}.

\subsection{Preliminary Tests}

We do some early geometric tests to confirm that the experiment is gives reasonable results, as a prelude to future work investigating particle isotropy and the physics of particle transport from the sun and the trapped electron model.\\

Rather than travelling in straight lines, the particles are deflected towards the poles and the South Atlantic Anomaly (SAA) by the earth's magnetic field, which is why the particle count is higher as LUCID gets towards the poles. We found slight evidence for marginally higher measured number counts when LUCID is in front of the earth than behind the earth (relative to the Sun), mean of 7.1 particles per frame `dayside'€™, and 6.3 particles `nightside'. Dayside and nightside were defined by calculating the angle made by the LUCID satellite, the centre of the earth and the sun using the latitude, longitude and timestamp of each frame. When this angle was between 0$^\circ$ and 90$^\circ$, the frame was classified as on the `dayside'€™ (i.e. the half of the earth facing the sun) and when it was more than 90$^\circ$ it was classified as on the `nightside'€™. Future work will focus on investigating if this is a systematic in the data or not, and testing our results within current models of the Earth's magnetosphere, as the Earth's magnetosphere is known to be different on the `day'€™ and `night'€™ sides of the Earth. Pressure from the solar wind coming from the sun on the dayside compresses the magnetic field, and on the nightside elongates it. In the Van Allen radiation belts (which LUCID is close to/passes through in the SAA), charged particles are trapped within the magnetic field, and where this field is condensed on the dayside there will be a higher particle density, which is consistent with a higher particle count on this side of the earth (e.g. \citealp{Williams1965NightsideKilometers}, \citealp{Domrachev2002ModelingMagnetosphere}, \citealp{Khazanov2002TransportMagnetosphere}). These results show that the LUCID data can contribute to testing and update models of space weather.

\section{Discussion} \label{sec:discussion}

In this section we discuss how our results compare to comparable space-based space radiation detectors, what our results mean for future use of Medipix in space, and what our future plans with LUCID data are.
 
\subsection{Other Experiments}

Our electron/proton percentages appear to be similar to other experiments. Our radiation maps are similar to those found by the ISS-REMs (e.g. Fig 7. \citealp{Stoffle2015Timepix-basedStation}). We do however go to higher latitudes (the ISS data only goes up to ~55$^\circ$N/S) mapping out the polar regions. SATRAM, \citet{Granja2016TheOrbit} also map to the higher altitudes. \citet{Litvak2017MonitoringExperiment} find similar plots in the neutron flux (measurements from the BTN-Neutron space experiment on the ISS), one of the particles LUCID can'€™t detect.\\

The ability of Timepix to classify heavy charged particles is highlighted in \citet{Kroupa2015ADosimetry}. Our results add to their findings that Timepix have potential for use as a `heavy ion'€™ monitor, or even telescope (\citealp{Branchesi2016Multi-messengerRays}). Furthermore, our use of human classifiers provides a testbed for future applications where humans are used to classify tracks that a traditional computer algorithm may struggle with.

\subsection{Future Plans}

Future work will focus on developing the algorithms and analyses presented in this work. In this paper we focussed on the classification of track morphologies. A subsequent paper will explore the calibration of energy measurements, and make dose maps, measurements of LET, and particle energy spectra. We will also redo our analysis with larger training sets and improved implementations of the deep learning approaches. This new analysis will use particle tracks that have been validated and calibrated from a defined radiation field so we can identify events more accurately.\\

We will investigate differences in flux measured in the different detectors (e.g. differences between Timepix perpendicular to each other). In addition, there is still much to exploit from the ability of Medipix detectors to measure both the elevation and azimuthal angle of particle trajectories in space e.g \citet{George2014MappingInformation}, \citet{Kroupa2015ADosimetry} find that particles are highly directional in the SAA (c.f. \citealp{Granja2014CharacterizationSATRAM}) See also rocket experiments at lower altitudes e.g. \citet{Zabori2016}. This would require identification of particle track angle and inclination as part of the machine learning classification algorithm, although we note that this must be done carefully to avoid uncertainty regarding the angles of entry.
In addition it would also be worthwhile to compare the student researcher track labels to labelled tracks from known sources - as already mentioned, the classifications of \citet{Vilalta2011DevelopmentDevices} and \citet{Hoang2013ADETECTORS} have the advantage of `true'€ labels compared with our human-classified labels, but may find it more difficult to identify unexpected tracks.\\

With classifications of particle inclination as well as particle type we will use the particle angle measurements to test the isotropy of particle flux in the trapped electron model. We will also investigate evolution in flux over time, to link measurements to solar cycles (\citealp{Thomas2014TheModulation}), and investigate the growth of the SAA and any possible link to geomagnetic reversal (\citealp{Pavon-Carrasco2016TheReversal}).\\

\added{It has recently been shown that the kinetic energy of particles can be reconstructed using a Timepix detector (\citealp{Kroupa2018}), which will also be a topic of future investigation with the information from the data that we have obtained from LUCID.}\
\

\subsection{Lessons From CERN@school}

We view LUCID as technological test of Medipix detectors for future space missions, giving Timepix data at a different altitude to other experiments, for the first time on a commercial platform, in a novel configuration of chips. However we also view CERN@school as an important test of the use of Medipix technology. Medipix had previously been predominantly used by professional scientists. CERN@school was the first widespread use of Medipix across 100s of institutions, with users ranging from novices to experts. The challenges of handling data from a very heterogeneous set of sources for use by a wide variety of users of different levels of expertise is a test bed for any future hypothetical large-scale use of Medipix outside of academia and research labs e.g. Medipix as a personal radiation monitor in a nuclear power plant or for nuclear medicine workers (\citealp{Michel2009LowMedipix}).\\

Alongside the research applications, CERN@school detectors in schools also have an educational role (e.g. as a replacement for a geiger counter in teaching about different types of radiation and the inverse square law). IRIS is currently undertaking pedagogical research about the effectiveness of this approach to teaching, and is developing plans to expand CERN@school to other CERN member states.

\section{Conclusions} \label{sec:conclusions}

In this paper we introduce the LUCID detector (the third use of Medipix detectors in space, the second in open space, and the first on a commercial platform and with a 3D configuration) that flew aboard TDS-1 taking data from 2014 to 2017. We describe the data pipeline from data collection to reduced catalogue of classified particles, a novel machine learning particle track classification algorithm, and some early science results. We also discuss LUCIDs links to the larger CERN@school ecosystem.\\

The payload was operated by submitting Payload Task Requests to SSTL. Data immediately after collection was stored on a flash partition on the spacecraft until passes over ground stations. The data was passed from the ground station to SSTL, and then to IRIS servers. From there the LUCID frame data were passed to GridPP for processing, and the reduced data passed back to the IRIS servers.\\

We use machine learning techniques to classify Timepix particle tracks, presenting two types of algorithm: a) a metric-based neural network classifier that uses eight pre-identified `features' and b) a deep learning approach, showing both perform well. These algorithms perform competitively compared with existing analytic algorithms used on large sets of Medipix data.\\

We present TAPAS, a new platform for Medipix/Timepix data that has been used extensively by secondary school researchers in the UK for managing large amounts of radiological data from a wide range of sources. TAPAS permits easy storage, sharing and visualisation via a browser interface.\\

Finally we discuss some early results from the reduced LUCID data. We find that the majority of particles detected are electrons, in agreement with the trapped electron model and earlier simulations of LUCID. Future work will focus on using LUCID data to characterise the radiation environment of LEO (in particular using angular information), and contribute towards planned future use of Medipix on  space missions.\\
 
Key results:
\begin{itemize}
\item We present the first results from LUCID, a novel space radiation detector that use Timepix detectors in orbit on TechDemoSat-1
\item We have developed the TAPAS platform, a versatile tool for managing Timepix datasets from a wide range of sources. We used TAPAS to handle large amounts of LUCID and CERN@school data, laying the groundwork for more widespread adoption of Timepix
\item We present the use of machine learning algorithms to classify track morphology and thus particle type in Medipix data, allowing for quick, accurate particle classification for huge amounts of data
\item Our results add to the evidence that Medipix detectors are well suited as space weather monitors/detectors
\item We have made preliminary flux maps (showing the SAA and similar features) at an altitude of ~600km, which appear similar to those of other detectors and present several other early science results, including detecting heavier charged particles
\item CERN@school has acted as a large scale trial of the use of Medipix in diverse environments by sizeable numbers of users 
\end{itemize}

\section*{Acknowledgements}

The authors and IRIS are extremely grateful to the Medipix Collaboration and SSTL for ten years of support, in particular we are hugely thankful to Dr Michael Campbell (CERN/Medipix Collaboration), Prof. Larry Pinksy (University of Houston/NASA), David Cooke (SSTL), Dr Stuart Eves (SSTL) and Dr Jonathan Eastwood (Imperial). Specific thanks to Prof. Stanislav Pospisil and IEAP CTU Prague for their support, in particular for calibrating the Timepix detectors used in LUCID.\\

Alongside the authors, the following students have also contributed substantially over the history of the project: Toby Freeland, Matt Harrison, Cal Hewitt, Sam Kittle, Nick Liu, Rachel O'€™Leary, Rachel Powell, Adam Sandey, Hector Stalker, Tom Stevenson and Cassie Warren, as well as extensive support and encouragement over many years from Dr Tom Whyntie (Oxford). Thank you also to students who contributed to the classifications used in the training set for the machine learning classifications, and the many other students who worked on LUCID over the last ten years.\\

A huge personal thank you from the authors to Prof. Becky Parker (IRIS) and Laura Thomas (IRIS) who have made LUCID and CERN@school happen, and have inspired countless students across the country over many years.\\

AS and EF would like to thank Mr Rupert Champion (Langton Star Centre) and Dr Tim Lesworth (Langton Star Centre) for providing ongoing support with their IRIS activities.
PH acknowledges funding from the Engineering and Physical Sciences Research Council.\\

The authors would like to thank Prof. Steven Rose (Imperial/Oxford) and Dr Carlos Granja (Advacam, Prague) for their helpful comments, remarks and input to the paper.\\

The authors and the Institute for Research in Schools would like to acknowledge generous support from: Humphrey Battcock, The Science and Technology Facilities Council, The Science Museum, The Institute of Physics, The Royal Commission for the Exhibition of 1851, The Ogden Trust, CERN, The Medipix Collaboration, NASA, the UK Space Agency, Kent County Council, IEAP CTU Prague, SEEDA and the GridPP Collaboration (in particular Dr Dan Traynor and Queen Mary University of London for use of their GPU and storage resources, and Professor Steve Lloyd for supporting and enabling schools to work with GridPP).\\

The authors would also like to gratefully thank the anonymous referees whose comments greatly improved the quality of this paper.\\

\appendix
\section[]{Code Access} \label{sec:code_access}

All the code discussed in this work is open source and available online:
\url{https://github.com/amshenoy/lucid_neural_analysis}\\
\url{https://github.com/willfurnell/lucid-grid/}\\
\url{https://github.com/InstituteForResearchInSchools/lucid_utils}\\

Information about LUCID, some frames from the mission, and some visualisation tools can be found at: \url{http://starserver.researchinschools.org/lucid_dashboard}

To support and assist the production of labelled LUCID data, the training web application can be found at: \url{http://starserver.researchinschools.org/lucid_trainer}

The TAPAS platform can be found at \url{https://tapas.researchinschools.org/}.
IRIS can be contacted for inquiries about accessing LUCID data via TAPAS at the supplied email address.

\section[]{Feature Definitions} \label{sec:appendix_feature_definitions}
This appendix gives the definitions of the features  used in the `Metric-Based Network'€ algorithm in Section \ref{sec:data_analysis}, and the details of the analytic classifier.

\begin{itemize}
  \item \textit{Find Average Neighbours} - For each individual pixel in the cluster, the number of neighbouring pixels is counted by iterating through the surrounding pixel coordinates and checking if they exist in the cluster. The pixels that exist are then stored in a list as these are the ones that have an energy deposit. The mean of the items in the list is then calculated and the function returns the average number of neighbours.
  \item \textit{Find Centroid (Centroid)} - The centroid of the cluster is the centre of the particle track. This centroid is found by calculating the mean of the x values and the mean of the y values of the pixels in the cluster. The mean x-value is the presumed to be the x-coordinate of the centre point and the mean y-value is presumed to be the y-coordinate of the centre point.  This coordinate is not the centre of the smallest enclosing circle but in fact, it is the centre of mass with no weighting on energy values. This function is used to calculate an auxiliary metric as the centroid is used to generate other metrics but it is not explicitly used for classification as it does not have any informative value by itself.
  \item  \textit{Calculate Radius (Radius \& Diameter)} - 

$\mathrm{Radius} = \sqrt{(x_c-x_n)^2+(y_c-y_n)^2}$

The Euclidean distance between the coordinates of each pixel to the centroid is calculated. The radius is set to the highest distance from the centroid. The diameter is then simply twice the calculated radius.

  \item \textit{Calculate Density (Density)} -

$\mathrm{Density}= \frac{\mathrm{Pixel Count}}{\pi r^{2}}$

The density can be greater than 1 as the cluster's radius passes through the centre of the outer pixels rather than around them. If the cluster is only one pixel in size, then the density is by default set to 1.

  \item  \textit{Calculate Non-Linearity (Line Residual)} - This function returns the angle $\theta$ anticlockwise from the x-axis, with the line passing through the cluster centroid. First, all the coordinates of the pixels within the cluster are split into separate lists of $X$ and $Y$ values. Single pixel tracks are by default given an angle and line residual of 0 as these are completely linear. For all other clusters, the above least squares regression function is used to calculate the line of best fit in the form . From the line of best fit, the best-fit angle  (anti-clockwise from the x-axis) of the line-of-best-fit can be calculated using simple trigonometry.
  
$b = \frac{\Sigma XY - \frac{\Sigma X \Sigma Y}{n} }{ \Sigma X^2 - \frac{(\Sigma X)^2}{n} }$\\

$a=Y-bX, \: \mathrm{   where   } \: X =\frac{\Sigma x}{n} \: \mathrm{and} \: Y =\frac{\Sigma y}{n}$\\

$\textrm{Line Residual} = \sum_{n=0}^{\mathrm{Pixel Count}} (x_{n}\sin \theta- y_{n}\cos \theta- x_{c}\sin \theta + y_{c}\cos \theta )^2 $\\

This best fit angle is used to calculate the line residual value using the equation above. The line residual value is calculated by the sum of the squares of the distance from each pixel coordinate to the line travelling through the centroid where the line is calculated using the best fit angle $\theta$. The equation above shows the simplest form of the line residual where  ($x_c$, $y_c$) is the coordinate of the centroid and ($x_n$, $y_n$) is the coordinate of the pixels in the cluster where $n$ is the index of the pixel in the cluster.

  \item  \textit{Find Best Fit Circle (Curvature Radius \& Circle Residual)} - This function is used to perform circle regression. For single pixel tracks, this cannot be done and therefore single pixels are by default given values of zero. As the cluster is a very poor initial guess for the centre of the circle, multiple test circles are generated using the calculated best fit angle from the regression line calculation. For each of these test circles, the program performs circle regression on the cluster using least squares. The Euclidean distance between the data points and the mean circle centred at ($x_c$, $y_c$) is calculated. The mean distance is then subtracted from all of the calculated distances. This value is optimised using the least squares function. The optimisation process returns an optimised centre point. This optimised centre point is used to calculate the distance of all the points from the centre of the circle once again.  The mean distance is set as the new radius and the residual is calculated by the summation of the squares of the difference between the distance from each point to the optimised centre point and the mean radius.

$(\textrm{Mean Radius}) \quad  R_{\mu}=\frac{\sqrt{(x_c-x_n)^2 + (y_c-y_n)^2}}{n}$\\

$\textrm{Circle Residual} = \Sigma_{n=0}^{\mathrm{Pixel Count}}(\sqrt{(x_c-x_n)^2 + (y_c-y_n)^2} - R_{\mu})^2 $\\

The optimised test circles are then organised in order of the magnitude of the residual. The aim of the optimisation is to reduce the circle residual as much as possible and therefore the circle with the least residual is chosen as the best fit circle.

\end{itemize} 

\textbf{Analytic Classifier} - The analytic classifier for particles used in early LUCID analysis is based on metrics that can be found in the open source \textit{lucid\_utils} library: \url{https://github.com/InstituteForResearchInSchools/lucid_utils/blob/master/lucid_utils/classification/lucid_algorithm.py}\\

\clearpage

\textbf{REFERENCES}

\bibliographystyle{elsarticle-harv_lucid}
\bibliography{lucid_asr_format}

\end{document}